\documentclass[twocolumn,showpacs,preprintnumbers,amsmath]{revtex4}

\usepackage{amssymb}
\usepackage{graphicx}
\usepackage{dcolumn}
\usepackage{bm}

\begin{document}

\title{Evidence for contact delocalization in atomic scale friction}

\author{D. Abel}

\author{S. Yu. Krylov}

\altaffiliation{Permanent address: Institute of Physical Chemistry, Russian Academy of Sciences, Leninsky
prospect 31, 119991 Moscow, Russia; e-mail: krylov@redline.ru}

\author{ J. W. M. Frenken}

\affiliation{Kamerlingh Onnes Laboratory, Leiden University, 2300 RA Leiden, The Netherlands}

\date{\today}

\begin{abstract}
We analyze an advanced two-spring model with an ultra-low effective tip mass to predict nontrivial and
physically rich "fine structure" in the atomic stick-slip motion in Friction Force Microscopy (FFM)
experiments. We demonstrate that this fine structure is present in recent, puzzling experiments. This shows
that the tip apex can be completely or partially delocalized, thus shedding new light on what is measured
in FFM and, possibly, what can happen with the asperities that establish the contact between macroscopic
sliding bodies.

\end{abstract}


\maketitle

%\preprint{APS/123-QED}


In the last two decades the Friction Force Microscope \cite{Mate} has become an essential tool for
nanotribology. In FFM experiments an atomically sharp tip is dragged along a surface by an external spring
(the cantilever), similar to AFM, and the lateral force is recorded with nN or even pN sensitivity. The FFM
tip is believed \cite{Salmeron} to model the behavior of a single asperity, similar to one of the many
asperities that make up the contact between two macroscopic sliding bodies, and thus provide direct,
atomic-scale access to the origin of friction. Often, FFM experiments demonstrate a periodic stick-slip
behavior of the lateral force, with the period of the substrate lattice. The FFM tip is thought to be held
periodically in a substrate lattice position until the increasing external force becomes sufficient to
force the tip to slip to the neighboring lattice position, etc.

FFM tips are not as rigid as they may seem at first glance; they are softer than most cantilevers
\cite{Gnecco,Dienwiebel,Socoliuc}. This inherent feature has long been believed not to complicate the
stick-slip physics. Traditionally, FFM is described by a single-spring (Tomlinson) model
\cite{Gnecco,Muser}, in which an effective mass, close to that of the cantilever, is dragged along the
surface by an effective spring, which accounts for the flexibility of both the cantilever and the tip.
First experimental and theoretical indications for the failure of the one-spring approach have appeared
only recently \cite{PRE2005,MaierPRB2005,ReimannPRL2004,PRL2006,EvstigneevPRL2006,Leng,Johnson}. For a true
understanding of the dynamics, we must explore at least a two-mass-two-spring scheme, one real mass $(M)$
accounting for the combined cantilever+tip inertia, and the other---effective---mass $(m)$ associated with
bending motion of the tip. This can introduce a wealth of new dynamics.

In ref. \cite{PRL2006} we have shown that a two-mass-two-spring system with a soft cantilever and
sufficiently low surface corrugation can exhibit strongly counterintuitive behavior of being ''stuck in
slipperiness'': the cantilever shows seemingly usual atomic stick-slip, while the tip-surface contact is
completely delocalized due to rapid, thermally activated motion of the tip apex back and forth between the
surface potential wells. An essential ingredient of this scenario is the assumption that the effective mass
$m$ is so low that the rate of thermally activated jumps of the tip apex is high with respect to the
characteristic frequency of motion of the cantilever-tip combination as a whole. This assumption was based
on the estimate \cite{PRL2006} that the bending deformation of an atomically sharp conical or pyramidal tip
is associated with only a few hundred atomic layers at its apex, so that $m\sim 10^{-20}$ kg, while the
typical value of $M$ is 9 to 12 orders of magnitude higher. With such an extreme mass ratio, the
low-frequency response of the cantilever, as measured in FFM, merely reflects the average of an ultra-fast
dynamics of the tip apex, which is actually probing the surface. Consequently, one can anticipate serious
changes in the description of atomic scale friction also in more typical cases of a hard cantilever and
higher surface corrugations. Unfortunately, a full Langevin description of the two-mass-two-spring system
\cite{MaierPRB2005} is not realistic for $m \lll M$ for computational reasons.

In this paper we report a theoretical analysis of the dynamics of a two-mass-two-spring system, which
provides a natural explanation of the peculiar "fine structure" of slip events recently observed in a
typical system with a hard cantilever \cite {MaierPRB2005}. The excellent agreement between theory and
experiment shows that the effective mass $m$ is ultra-low indeed and the nanocontact can be completely or
partially delocalized. This sheds new light not only on what is actually measured in FFM, but also what can
possibly happen with the asperities that establish the contact between two macroscopic sliding bodies.

We have developed a hybrid computational scheme that combines a numerical Langevin description of the
cantilever+tip motion with a Monte-Carlo treatment of the thermally activated motion of the tip's apex. It
enables one to follow the dynamic interplay between the rapid motion of $m$ and the slow motion of $M.$ We
find a surprisingly wide variety of dynamic behavior, depending sensitively on the masses, spring constants
and the surface corrugation. The rapid transition dynamics of $m$ is not washed out in the slow response of
$M,$ but leads to the existence of several different observable regimes, including situations with a
delocalized tip-surface contact. The variety of regimes will be discussed in detail elsewhere. Here we
concentrate on a particular but very important issue which is most closely related with the origin of
dissipation and allows critical comparison with experiment. Our calculations show that the slipping of the
cantilever---the processes when energy stored in the system is rapidly dissipating---can proceed in several
different ways, depending sensitively on the surface corrugation. Besides the ''fast'' slipping, there are
slip events that take more time and have nontrivial ''fine structure''. These unusual slips directly
reflect delocalization of the tip-surface contact. Our results find remarkably good confirmation in the
recent observations \cite{MaierPRB2005}, a high-resolution experiment in which durations of slip events
have been resolved for the first time. This provides a straightforward explanation for the unique
experimental work, the authors of which have ascribed the unusual slip events to a possible but highly
improbable configuration of simultaneous contact via several ''nanotips'', positioned commensurate with the
substrate lattice.

For a one-dimensional geometry, the total potential energy of the system can be written as
\begin{equation}
U(X,x,t)=\frac{K}{2}(Vt-X)^{2}+\frac{k}{2}(X-x)^{2}+U_{\text{s}}(x)\,, \label{Utot}
\end{equation}
with $X$ and $x$ the coordinates of the cantilever and the tip apex, respectively; $Vt$ is the position of
the support that moves with the scanning velocity $V;$ $K$ and $k$ denote the stiffness of the cantilever
and of the tip. The tip--surface interaction is assumed to be sinusoidal, with amplitude $U_{0}$ and period
$a,$ $U_{\text{s}}=\frac{U_{o} }{2}[1-\cos (\frac{2\pi x}{a})].$ The system is described by two coupled
equations of motion, one for the cantilever+tip combination (position $X$ and mass $M$) and the other for
the tip apex (position $x$ and effective mass $m)$ moving with respect to $X$. If $m\lll M,$ and hence
there is a strong hierarchy between the characteristic frequencies of the tip apex $(\nu _{\text{t}})$ and
the cantilever $(\nu _{\text{c} }),$ $\nu _{\text{t}}\gg \nu _{\text{c}},$ the description can be
simplified by averaging over the rapid thermal motion of the apex around lattice positions $x_{i}.$ For
each position of the cantilever $X$, the $x_{i}(X)$ correspond to the local minima in the total potential
(\ref{Utot}) as a function of $x.$ The number of wells available to the apex is determined by the Tomlinson
parameter $\gamma =\frac{2\pi ^{2}U_{0}}{ka^{2}}$ \cite {PRL2006}. If $\gamma
>1,$ there are two or more wells. Not only is this the origin of stick-slip motion, this also introduces
the possibility of thermally activated jumps of the tip apex between the wells. Here we restrict ourselves
to the simplest (transition state theory) approximation to the jump rate: $r_{ij}=\nu _{\text{t}}\exp
\left( -\frac{U_{ij}}{k_{B}T} \right) ,$ with $U_{ij}(X)$ the potential barrier between wells $i$ and $j$.
Following this scheme, one can describe motion of the cantilever by solving numerically only a single
Langevin-type equation,
\begin{equation}
M\stackrel{\cdot \cdot }{X}=-k\left[ X-x_{i}(X)\right] -K(X-Vt)-M\eta \stackrel{\cdot }{X}+\xi \;,
\label{M}
\end{equation}
in combination with a Monte-Carlo algorithm for transitions of the tip apex between positions $x_{i}$ and
$x_{j}$ with rate $r_{ij}.$ The random force $\xi $ is normalized as $\langle\xi (t)\xi (t^{\prime
})\rangle=2M\eta _{\text{n}}k_{\text{B} }T\delta (t-t^{\prime }).$ According to the fluctuation-dissipation
theorem for a particle interacting with a bath, $\eta _{\text{n}}=\eta .$ In our case the cantilever is
coupled to the bath very indirectly, via motion of the tip apex with respect to the cantilever and with
respect to the surface (damping in the cantilever can be neglected \cite{MaierPRB2005}). In order to
control the possible role of damping and noise, we varied both $\eta $ and $\eta _{\text{ n}}$ in our
calculations. The results presented below correspond to the case of slightly overdamped motion, $\eta =5\nu
_{\text{c}},$ while the noise has been artificially reduced (by a factor of $\sqrt{10})$ by taking $\eta _{
\text{n}}=0.5\nu _{\text{c}}$ in order to better visualize the fine structure of the slip events. We
checked that in a wide range $0.1\,\nu _{\text{c}}<\eta <10\,\nu _{\text{c}}$ the results do not change
qualitatively although they contain stronger fluctuations at lower damping and higher random force
amplitude.

The results in Fig. 1 have been obtained assuming $m=1\cdot 10^{-20}$ kg, as estimated in \cite{PRL2006},
and with all other parameters taken from the experiment of Ref. \cite{MaierPRB2005}: $a=0.66$ nm, $K=62$
N/m, $V=25$ nm/s, $T=300$ K and $M=5.5\cdot 10^{-11}$ kg. The value of the tip stiffness $k=3$ N/m chosen
here from the best fit to the experimental stick-slip patterns at high corrugations (Fig. 2-a of
Ref.\cite{MaierPRB2005}) is in the range of several N/m typical for FFM tips. Varying the surface
corrugation $U_{0}$, corresponding to changing scan lines in the experiment, we pass through different
regimes: from stick-slip with trivial slips at high $U_{0},$ via several regimes with ''structured'' slips
at lower corrugations, to near-dissipationless motion \cite{Dienwiebel,Socoliuc} at very low corrugation
$(\gamma <1).$ Representative examples of slip events in different regimes are shown in Fig. 1 (d-i).

At high surface corrugation, when the probability of thermally activated jumps of the tip apex is low, each
slip to the next accessible well is followed by the corresponding slip of the whole cantilever (Fig. 1-d).
The latter takes place on a time scale of $\nu _{\text{c}}^{-1}$ . At somewhat lower corrugation the jump
rate is larger, so that the apex exhibits several additional jumps back and forth between the wells, and so
does the cantilever (Fig. 1-e,f). (Note that the behavior of the apex is directly accessible in the
calculations but not shown here.) At lower $U_{0},$ the cantilever cannot follow the rapid jumps of the tip
apex, so it exhibits a smoothened stochastic behavior (Fig. 1-g).

The physics behind the gradual slips (Fig. 1-h) and the slips with an intermediate state (Fig. 1-i) is more
complex. For the parameter values used in the calculations, $\gamma $ is not too far from unity, so that
the tip apex experiences a double-well potential when the cantilever is close to midway between lattice
sites $(X=a/2,$ $3a/2,$ $...),$ otherwise it sees only one well. The potential barrier between the two
wells is considerably lower than the surface corrugation $U_{0},$ so that the rate of thermally activated
jumps can be very high. In the example of Fig. 1-i, the mean jump rate exceeds $0.5\cdot 10^{9}$ Hz over
nearly the entire $X$-interval with two wells, thus being high above the characteristic frequency $\nu
_{\text{c}}$ of the cantilever. The tip apex is delocalized due to rapid jumps back and forth between the
two accessible wells, while the cantilever sees only the mean tip apex position. In other words, the
cantilever moves in an effective potential formed by the rapid thermal motion of the tip apex. The
effective tip--surface interaction can be calculated as
$U_{\text{eff}}(X)=\int\nolimits_{0}^{X}\overline{F}_{\text{s} }(X^{\prime })dX^{\prime },$ with the mean
force exerted on the cantilever by the bending tip $\overline{F}_{\text{s}}(X)=-k(X-\overline{x}).$ The
mean position of the apex is $\overline{x}(X)=x_{1}p_{1}+x_{2}p_{2},$ with $ p_{1}$ and $p_{2}$ the
probabilities to find it in well 1 or 2 respectively. From the equilibrium distribution of apex positions,
a good approximation for the case $V/a \ll \nu _{\text{c} } \ll \overline{r}$ considered, one easily finds
$p_{1}=r_{21}/(r_{12}+r_{21}),$ $p_{2}=r_{12}/(r_{12}+r_{21}).$ $U_{\text{eff} }(X)$ and
$\overline{F}_{\text{s}}(X)$ are plotted in Fig. 2 for the parameters used in Figs. 1-h,i. The mean surface
force is seen to exhibit discontinuities at the positions of the cantilever where the system experiences
''sudden'' transitions of the tip apex potential from a one-well to a two-well shape, and vice versa.
Actually, the transitions must be smooth over a time of the order of $\overline{r}$ $^{-1}$ needed for
establishing the equilibrium population of the two wells involved. Within the two-well interval, the mean
force is seen to be essentially reduced, while the effective potential is flattened at the top and even can
have a shallow minimum (dotted line in Fig. 2-a). The reduction of the mean surface force can be understood
as a result of the rapid oscillations of the instantaneous force: sometimes it is positive, sometimes
negative. The corresponding flattening of the effective potential can be viewed as a result of the
competition between two trends: when the the cantilever approaches the maximum of the surface corrugation,
the mean potential energy of the system increases, but also its entropy increases by disordering in
positions of the apex. In this way one understands, that the gradual slip of the tip seen in Fig. 1-h
reflects a relatively gradual change in the mean surface force (dashed line in Fig. 2-b). If the change is
more step-like (solid line), the cantilever exhibits slip via an intermediate state (Fig. 1-i), which
reflects a nearly zero mean surface force acting in this interval of the cantilever positions.

Interestingly, conditions under which the cantilever exhibits only a small number of jumps per lattice
spacing, e.g. single slips, are found not only at high $U_{0}$ but also at very low surface corrugations,
when $\gamma $ is very close to unity (Fig. 1-k,l). In this case the interval of cantilever positions for
which the apex sees two surface wells is so narrow that it is rapidly passed with hardly any response to
the temporarily rapid jump dynamics of the apex.

Our predictions find direct confirmation in recent observations Ref. \cite{MaierPRB2005}, reproduced in
Fig. 3. Compare Fig. 1-d,l with Fig. 3-b; Fig. 1-e,k with Fig. 3-c; Fig. 1-h with Fig. 3-a; and, most
importantly, Fig. 1-i with Fig. 3-d. Besides one-to-one reproduction of the types of fine structure in slip
events, there is also remarkably good correspondence between the measured and simulated ranges of variation
of the lateral force. This is in spite of the strong simplifications concerned with the reduced
dimensionality and the simple sinusoidal potential assumed in our calculations. The only discrepancy
concerns the systematic difference in the durations of all the structured slips: in the experiment they are
larger than in our simulations. This suggests that the actual surface corrugation is not simply sinusoidal,
but has a specific shape characterized by wider barriers and narrower wells. Potential landscapes of this
type have been encountered before \cite{notsine}. As is clear from the discussion above, such a modified
surface potential should indeed be accompanied by an increased slip duration. Moreover, it should lead to
an increase in the maximum lateral force for the same corrugation, thus also improving the correspondence
between theory and experiment in this respect.

From simulations for different values of $m,$ we obtain evidence that the actual effective mass of the
contact is very small indeed. A good criterion for this is the occurrence of slips with an intermediate
state (Figs. 1-i and 3-d), the most specific slips, directly related with complete delocalization of the
contact in certain intervals of the cantilever positions. Decreasing $m$ from $10^{-20}$ kg by three orders
of magnitude, we do not see sizable changes in the results. However, if $m$ is increased by two orders of
magnitude, the slips with intermediate state become less pronounced and shorter in time; after a further
order-of-magnitude increase of $m$ (still $m\ll M)$ the intermediate state is not seen anymore. From this
we obtain the upper estimate $m<10^{-18}$ kg, which contrasts earlier expectations
\cite{ReimannPRL2004,MaierPRB2005} but agrees with our recent calculations \cite{PRL2006}.

The evidence presented here for a very small effective mass and related delocalization of the tip-surface
contact brings us to several important conclusions. First, there is a variety of different regimes of
friction, including the ''stuck-in-slipperiness'' predicted in \cite{PRL2006} for a system with a soft
cantilever. Note that in the latter case the contact can be completely delocalized at any position of
cantilever, rather than in limited intervals, as described above. Second, interpretation of the lateral
force amplitude, related to the corrugation of the tip-substrate interaction energy, the slip times and the
thermal noise in many seemingly standard FFM measurements will have to be reconsidered. Third, in different
regimes, like in cases (d-i) in Fig. 1, one meets essentially different scenarios of energy dissipation. As
suggested by our results, the key element is the rapid motion of the tip apex, with a characteristic
frequency in the GHz range \cite{PRL2006}. Apparently, it can readily relax due to two complimentary
mechanisms: damping of the intratip vibrations, accompanied by creation of phonons inside the tip, and
damping of the apex motion with respect to the substrate, accompanied by creation of phonons both in the
substrate and the tip. Which of these two energy sinks is more important remains unresolved yet.

Finally, the effects highlighted here for the case of an FFM tip in contact with a substrate can possibly
also play a role in the much more general context of the asperities that establish the contact between two
macroscopic sliding bodies. In other words, there may be much more thermally driven dynamics in macroscopic
sliding due to the local compliance of the contacting surfaces than we have ever imagined.

This work was supported by the Foundation for Fundamental Research on Matter (The Netherlands).

\newpage

\begin{center}
F I G U R E C A P T I O N S
\end{center}

\medskip

Fig. 1. Lateral force as a function of support position. Examples of atomic stick-slip at high and moderate
surface corrugation and continuous sliding at low corrugation: $U_0=1.2,$ (a) $0.48$ (b) and $0.3$ eV (c).
Representative examples of slip events (zoomed in) for different corrugations: 1.2 (d), 1.0 (e), 0.9 (f),
0.7 (g), 0.52 (h), 0.48 (i and j), 0.43 (k) and 0.42 eV (l). Calculations for $m=10^{-20}$ kg and all other
parameters taken from the experiment of Ref. \cite{MaierPRB2005}. In (j) thermal noise has been switched
off $(\eta _{\text{n}}=0)$ to better visualize the intermediate state accompanying the contact
delocalization.

Fig. 2. Effects of contact delocalization in the effective tip-surface interaction (a) and the mean surface
force (b) versus cantilever position, for $U_0=0.52$ (dashed lines), 0.48 (solid) and 0.44 eV (dotted). The
other parameters are as in Fig. 1.

Fig. 3. Examples of experimentally observed slip events (reproduced from Ref. \cite{MaierPRB2005}).

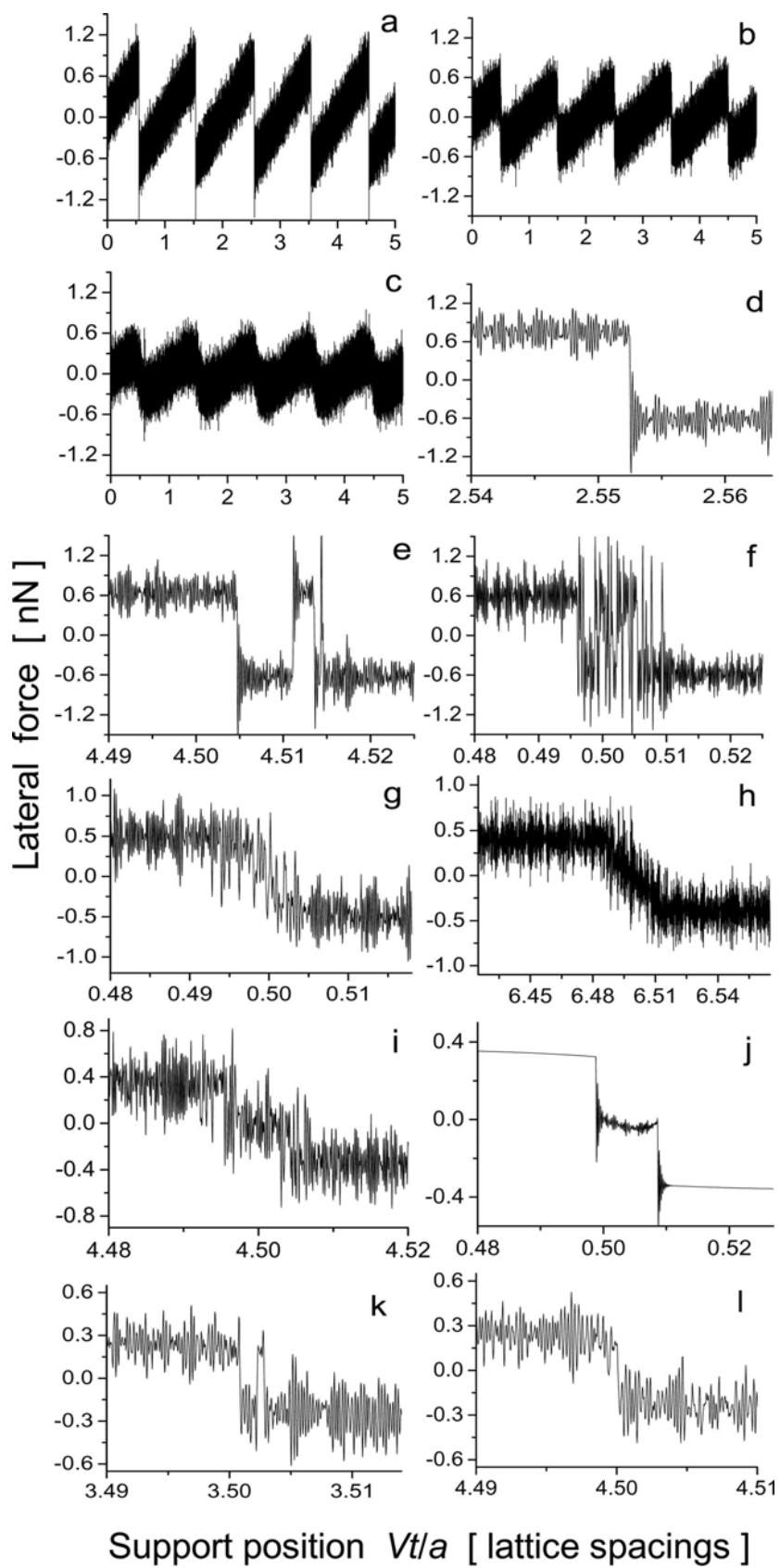
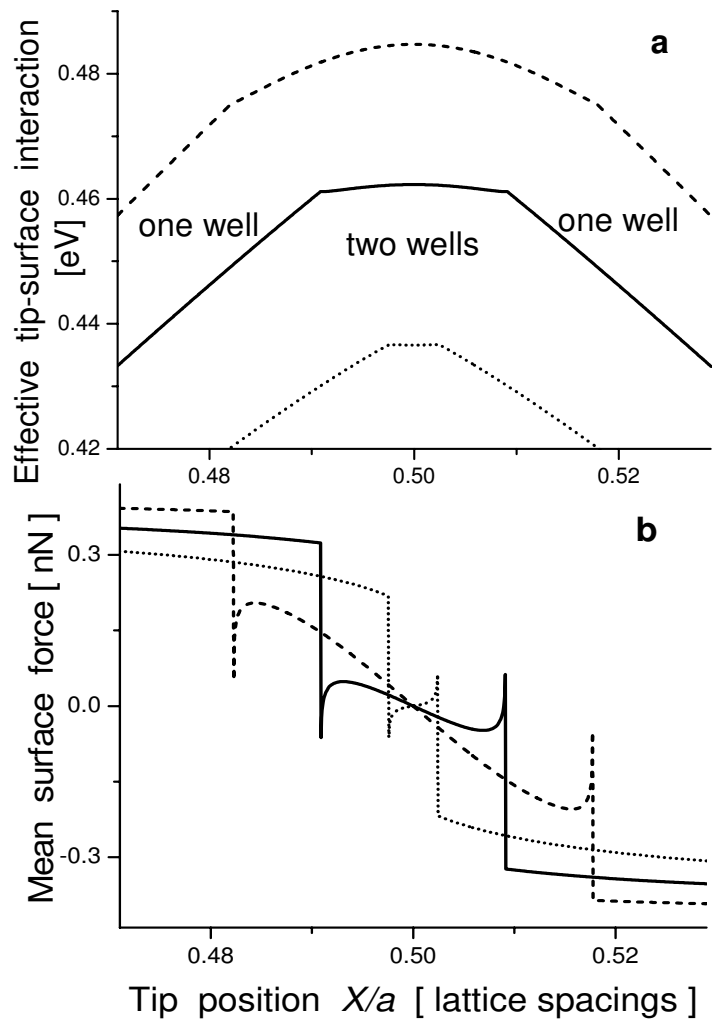
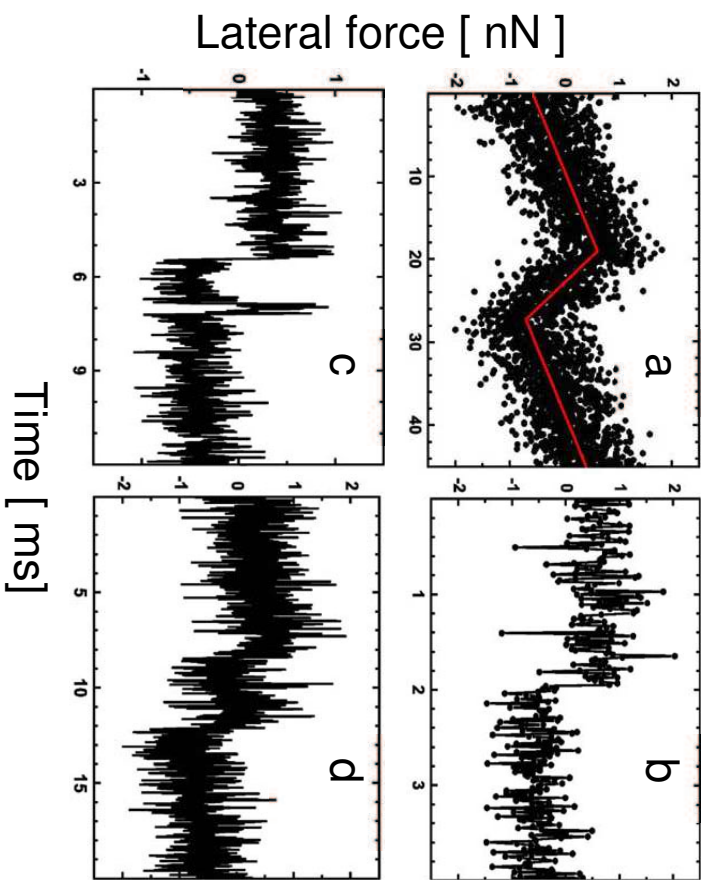
\end{document}